\documentclass[preprint,prd,aps,nofootinbib]{revtex4}
\usepackage{graphicx}
\usepackage{color}
\UseRawInputEncoding

\begin{document}


\title{W boson mass in the NP models with extra $U(1)$ gauge group}

\author{Jin-Lei Yang$^{1,2}$\footnote{jlyang@hbu.edu.cn},Zhao-Feng Ge$^{3}$\footnote{gezhaofeng@itp.ac.cn},Xiu-Yi Yang$^{4}$\footnote{yxyruxi@163.com},Sheng-Kai Cui$^{1,2}$,Tai-Fu Feng$^{1,2}$}

\affiliation{Department of Physics, Hebei University, Baoding, 071002, China$^1$\\
Key Laboratory of High-precision Computation and Application of Quantum Field Theory of Hebei Province, Baoding, 071002, China$^2$\\
CAS Key Laboratory of Theoretical Physics, School of Physical Sciences, University of Chinese Academy of Sciences, Beijing 100049, China$^3$\\
School of science, University of Science and Technology Liaoning, Anshan, 114051, China$^4$}

\begin{abstract}
The precise measurement of the W boson mass is closely related to the contributions of new physics (NP), which can significantly constrain the parameter space of NP models, particularly those with an additional $U(1)$ local gauge group. The inclusion of a new $Z'$ gauge boson and gauge couplings in these models can contribute to the oblique parameters $S$, $T$, $U$ and the W boson mass at tree level. Taking into account the effects of kinetic mixing, in this study we calculate and analyze the oblique parameters $S$, $T$, $U$ and the W boson mass in such NP models. It is found that the kinetic mixing effects can make significant contributions to the W boson mass, and these contributions can be eliminated by redefining the gauge boson fields through eliminating the neutral currents involving charged leptons, if the leptonic Yukawa couplings are invariant under the extra $U(1)$ local gauge group, even with nonzero kinetic mixing effects.
\end{abstract}

\maketitle

\section{Introduction\label{sec1}}

The imposing of gauge invariant in the Standard Model (SM) requires the introducing of gauge bosons. Due to the fact that the Lagrangian involving Higgs doublet in the SM is invariant under the $SU(2)_L\otimes U(1)_Y$ gauge transformation, the electroweak gauge bosons acquire nonzero masses through the spontaneous symmetry breaking (SSB). This SSB mechanism also provides mass to all massive particles in the SM. Therefore, high-precision measurements of gauge boson masses are crucial for testing SM and searching for NP phenomena. Recently, the CDF II measurement~\cite{CDF:2022hxs} of the W boson mass in 2022 is approximately $7\sigma$ away from the SM prediction~\cite{Haller:2022eyb}
\begin{eqnarray}
&&M_W^{\rm SM}=80354\pm7\;{\rm MeV},\\
&&M_W^{\rm CDF\;II}=80433.5\pm9.4\;{\rm MeV}.
\end{eqnarray}
And $M_W^{\rm CDF\;II}$ is not compatible with the world average
\begin{eqnarray}
&&M_W^{\rm Average}=80377\pm12\;{\rm MeV},
\end{eqnarray}
which is based on the measurements at LEP-2~\cite{ALEPH:2013dgf}, Tevatron~\cite{CDF:2012gpf,D0:2013jba} and the LHC~\cite{ATLAS:2017rzl,LHCb:2021bjt}. The measured W boson mass anomaly at CDF II has prompted numerous studies attempting to explain it within various NP models~\cite{Athron:2022uzz,Lu:2022bgw,DAlise:2022ypp,Strumia:2022qkt,Zhu:2022tpr,Athron:2022qpo,deBlas:2022hdk,FileviezPerez:2022lxp,Athron:2022isz,Babu:2022pdn,Bahl:2022xzi,
Sakurai:2022hwh,Nagao:2022dgl,Du:2022brr,Zeng:2022lkk,Barger:2022wih,Primulando:2022vip,Kawamura:2022fhm,VanLoi:2022eir,Cao:2022mif,Heckman:2022the,Alguero:2022est,
Allanach:2022bik,Senjanovic:2022zwy,Mandal:2023oyh}. In a recent update, the ATLAS collaboration reported a measurement of the W boson mass based on data from proton-proton (pp) collisions at a center-of-mass energy $\sqrt s=7\;{\rm TeV}$, the result reads~\cite{ATLAS:2023fsi}
\begin{eqnarray}
&&M_W^{\rm ATLAS}=80360\pm 16\;{\rm MeV}.
\end{eqnarray}
Interestingly, the updated result from the ATLAS collaboration shows agreement with the SM prediction, with no observed deviation from the SM expectation. This underscores the importance of obtaining a proper W boson mass that can satisfy experimental measurements in NP models, which is crucial for NP phenomenology studies.

Among the various NP models, those with an extra $U(1)$ gauge group, such as the $U(1)_{B-L}$ extended SM, the minimal supersymmetric model (MSSM) with local $B-L$ gauge symmetry~\cite{Khalil:2008ps,Basso:2012gz,Elsayed:2012ec,Khalil:2015naa,Yang:2019aao,Yang:2020ebs,Yang:2020bmh,Yang:2021duj}, the MSSM with $U(1)_X$ local gauge group~\cite{Bandyopadhyay:2011qm,Belanger:2011rs,Belanger:2015cra,Belanger:2017vpq,Zhao:2019ggv} and etc, have attracted the attention of physicists. In these NP models, the introduction of a new $Z'$ gauge boson corresponding the extra $U(1)$ gauge group and new introduced gauge couplings can make contributions to the oblique parameters $S$, $T$, $U$ at the tree level, and consequently impact the theoretical predictions on the W boson mass. Therefore, determining the appropriate gauge coupling strength and the new $U(1)$ gauge group charges of the particle fields that can satisfy the high-precision measurements of the W boson mass is crucial in these NP models. In this work, we present a general analysis of the W boson mass in the NP models with a new extra $U(1)$ gauge group, and the obtained results can be applied to all such NP models.

The paper is organized as follows. The $S$, $T$, $U$ parameters and W boson mass in the NP models with extra $U(1)$ local gauge group are calculated in Sec.~\ref{sec2}. The numerical results are presented and analyzed in Sec.~\ref{sec3}. Finally, a summary is made in Sec.~\ref{sec4}.

\section{W boson mass in the NP models with extra $U(1)$ local gauge group\label{sec2}}

For convenience, the newly introduced $U(1)$ local gauge group can be defined as $U(1)_X$ ($X=B,\;L,\;B-L$ and etc, where $B$ refers to the baryon number, $L$ refers to the lepton number), i.e. the local gauge group of this kind of NP models is extended to $SU(3)_C\otimes SU(2)_L\otimes U(1)_Y\otimes U(1)_X$. Due to the presence of two Abelian groups, the gauge kinetic mixing can occur and can be induced through RGEs even if it is set to zero at $M_{\rm GUT}$~\cite{Holdom:1985ag,Matsuoka:1986ig,delAguila:1987st,delAguila:1988jz,Foot:1991kb,Babu:1997st}. This leads to the covariant derivatives in this type of NP model being expressed as
\begin{eqnarray}
&&D_\mu=\partial_\mu-i\left(\begin{array}{c}Y,\;\;X\end{array}\right)\left(\begin{array}{c}g_Y,\;\;\;g_{YX}\\g_{XY},\;\;\;\;g_X\end{array}\right)
\left(\begin{array}{c}B_\mu\\ Z'_\mu\end{array}\right),
\end{eqnarray}
where $Y,\; X$ are the hypercharge and $U(1)_X$ charge respectively, $g_Y$ is the measured hypercharge coupling constant, $g_X$ is the coupling constant corresponding to $U(1)_X$ gauge group, $g_{YX}$ is the coupling constant arises from the gauge kinetic mixing effect, while $g_{XY}$ can always be rotated to $0$ as long as the two Abelian gauge groups are unbroken.

The kinetic terms of vector bosons $\Pi_{33}(p^2)$, $\Pi_{00}(p^2)$, $\Pi_{30}(p^2)$, $\Pi_{WW}(p^2)$ can be defined by the effective Lagrangian~\cite{Marandella:2005wd}
\begin{eqnarray}
&&\mathcal{L}_{\rm oblique}=-\frac{1}{2}W_\mu^3 \Pi_{33}(p^2)W^{3\mu}-\frac{1}{2}B_\mu \Pi_{00}(p^2)B^\mu-W_\mu^3 \Pi_{30}(p^2)B^\mu\nonumber\\
&&\qquad\qquad\;-W_\mu^+\Pi_{WW}(p^2)W^{-\mu}.\label{eq6}
\end{eqnarray}
$\Pi_{ij}(p^2)$ ($ij=33,\;00,\;30,\;WW$) in Eq.~(\ref{eq6}) can be expanded at $p^2=0$ because the NP is generally considered to be very heavy, and $\Pi_{ij}(p^2)$ can be written as
\begin{eqnarray}
&&\Pi_{ij}(p^2)=\Pi_{ij}(0)+p^2 \Pi'_{ij}(0)+\cdot\cdot\cdot,
\end{eqnarray}
where the higher order terms can be neglected safely. The oblique parameters $S,\;T,\;U$ are defined as
\begin{eqnarray}
&&S=4s_Wc_W\Pi'_{30}(0),\;T=\frac{\Pi_{33}(0)-\Pi_{WW}(0)}{M_W^2},\;U=4s_W^2[\Pi'_{33}(0)-\Pi'_{WW}(0)],\label{eq8}
\end{eqnarray}
where $s_W\equiv \sin \theta_W,\;c_W\equiv \cos \theta_W$, and $\theta_W$ is the Weinberg angle.

Generally, the $\Pi$ matrix of neutral vector bosons can be written in the basis $(B_\mu,\;W^3_\mu,\;Z'_\mu)$ as
\begin{eqnarray}
&&\left(\begin{array}{ccc}p^2-M_W^2 t^2 & M_W^2 t & -M_W^2 t t'\\
M_W^2 t & p^2-M_W^2 & M_W^2 t'\\
-M_W^2 t t' & M_W^2 t' & p^2-M_{Z'}^2\end{array}\right),\label{eq9}
\end{eqnarray}
where $t=g_{Y}/g_2$, $t'=(g_{YX}+2X_H g_X)/g_2$. It is obvious that there is nonzero $Z-Z'$ mixing effect even $X_H=0$ because of the existence of kinetic mixing effect. In general, the kinetic mixing effect can also be described by the kinetic mixing parameter $\epsilon$
\begin{eqnarray}
&&L=-\frac{1}{4}B_{\mu\nu}B^{\mu\nu}-\frac{1}{4}X_{\mu\nu}X^{\mu\nu}-\frac{\epsilon}{2}B_{\mu\nu}X^{\mu\nu}+\frac{1}{2}M_Z^2 B_\mu B^\mu+\frac{1}{2}M_{Z'}^2 X_\mu X^\mu.
\end{eqnarray}
The kinetic terms of the Lagrangian above can be normalized by
\begin{eqnarray}
&&K=\left(\begin{array}{cc}-k_1 & k_2\\
k_1 & k_2\end{array}\right),
\end{eqnarray}
where $k_1=1/\sqrt{1-2\epsilon}$, $k_2=1/\sqrt{1+2\epsilon}$. The normalization arises the nonzero mixing terms of the mass matrix for $(B_\mu,\;X_\mu)$
\begin{eqnarray}
&&\left(\begin{array}{cc}k_1^2 (M_{Z'}^2+M_Z^2) & k_1k_2 (M_{Z'}^2-M_Z^2)\\
k_1k_2 (M_{Z'}^2-M_Z^2) & k_2^2 (M_{Z'}^2+M_Z^2)\end{array}\right).\label{eqa12}
\end{eqnarray}
Combining Eq.~(\ref{eq9}) with Eq.~(\ref{eqa12}), we can obtain
\begin{eqnarray}
&&\epsilon\approx\frac{1}{2}-\frac{(g_{YX}+2X_H g_X)^2M_W^4}{2g_2^2M_{Z'}^4}.
\end{eqnarray}
Hence, it is equivalent to describe the kinetic mixing effect by Eq.~(\ref{eq9}) and Eq.~(\ref{eqa12}), we will take the forms defined in Eq.~(\ref{eq9}) to carry out the following analysis.

The gauge couplings involving charged leptons can be written as
\begin{eqnarray}
&&\mathcal{L}=-i[(-\frac{1}{2}g_2W^3_\mu+Y_L g_Y B_\mu+Y_L g_{YX} Z'_\mu+X_L g_B Z'_\mu)\bar e_L e_L\nonumber\\
&&\qquad+(g_Y Y_E B_\mu+g_B X_E Z'_\mu+g_{YX} Y_E Z'_\mu)\bar e_R e_R]\label{eq10}
\end{eqnarray}
$Y_L$, $Y_E$ are the hypercharges of the left-handed and right-handed components of leptons respectively, they can be normalized as $Y_L=-\frac{1}{2}$, $Y_E=-1$. And $X_L,\;X_E,\;X_H$ are the $U(1)_X$ charges of left-handed components of leptons, right-handed components of leptons and scalar doublets respectively. Since the most strict constraints come from the precision measurements performed at $e^+e^-$ colliders (such as LEP1, LEP2 and etc), we choose to eliminate the neutral currents involving charged leptons in Eq.~(\ref{eq10}), which can be done by redefining the vector fields
\begin{eqnarray}
&&B_\mu=\tilde B_\mu-c_Y Z'_\mu,\;\;W_\mu^3=\tilde W_\mu^3-c_W Z'_\mu,\nonumber\\
&&Z'_\mu=\tilde Z'_\mu+\frac{1}{p^2-M_{Z'}^2}\{[(p^2-M_W^2 t^2)A_Y+M_W^2 tA_W+M_W^2 tt']\tilde B_\mu+[M_W^2 tA_Y\nonumber\\
&&\qquad\;\;+(p^2-M_W^2)A_W-M_W^2 t']\tilde W^3_\mu\},\nonumber\\
&&A_Y=\frac{g_X X_E+g_{YX}Y_E}{g_1 Y_E},\;\;A_W=\frac{2}{g_2}(Y_L g_Y c_Y-X_L g_B-Y_L g_{YX})
\end{eqnarray}
Then the oblique parameters $S,\;T,\;U$ can be obtained from Eq.~(\ref{eq8}) as
\begin{eqnarray}
&&S=\frac{8 s_W^2 M_W^2 g_X}{M_{Z'}^2}[\frac{1}{g_Y^2}(g_X X_E-g_{YX})+\frac{g_X}{g_2^2}(X_E-2X_L)](X_E-X_L+X_H),\label{eq12}\\
&&T=\frac{4g_X^2M_W^2}{g_2^2 M_{Z'}^2}(X_E-X_L+X_H)^2,\label{eq13}\\
&&U=\frac{-8g_X^2M_W^2}{g_2^2 M_{Z'}^2}(X_E-2X_L)(X_E-X_L+X_H).\label{eq14}
\end{eqnarray}
As can be seen in Eqs.~(\ref{eq12}-\ref{eq14}) that $S=T=U=0$ when $X_E-X_L+X_H=0$, it indicates the leptonic Yukawa couplings are invariant under the $U(1)_X$ symmetry in this case which avoids the need of a model to generate the lepton masses. And for nonzero $X_E-X_L+X_H$, other mechanisms are needed to generate the lepton masses, such as the $U(1)_H$ model~\cite{Mira:1999fx} and etc. In addition, we neglect the $Z$ couplings to neutrinos or quarks in our calculations approximately, because they are measured much less accurately than $Z$ couplings to charged leptons, and all effects involving charged leptons are included in the approximation. The W boson mass with local $U(1)_X$ gauge group can be written as~\cite{Peskin:1990zt,Peskin:1991sw}
\begin{eqnarray}
&&M_W^2=(M_W^{\rm SM})^2+\frac{c_W^2 M_Z^2}{c_W^2-s_W^2}\Big(-\frac{1}{2}S+c_W^2 T+\frac{c_W^2-s_W^2}{4s_W^2}U\Big).
\end{eqnarray}

Eqs.~(\ref{eq12}-\ref{eq14}) indicate that the $U(1)_X$ charges of leptons, denoted as $X_E$ and $X_L$, can affect the theoretical predictions for the $W$ boson mass. Consequently, we will consider $X_E$ and $X_L$ as free parameters to explore their effects in the following analysis. Typically, $X_E$ and $X_L$ are subject to constraints enforced by chiral anomaly cancellations. For NP models with an extra $U(1)_X$ local gauge group, there are four distinctive chiral anomaly cancellation patterns for per generation~\cite{Nagao:2022dgl}
\begin{eqnarray}
&&\sum_{f}X_f=\sum_{f}(X_f)^3=\sum_{f}(X_f)^2Y_f=\sum_{f}X_f(Y_f)^2=0.\label{eq16}
\end{eqnarray}
Here, $f$ represents various fermion species, encompassing the left-handed lepton doublet $\hat L$, the left-handed quark doublet $\hat Q$, the right-handed lepton singlets $\hat E$ and $\hat N$ (heavy neutrinos), and the right-handed quark singlets $\hat u$ and $\hat d$. It is evident from Eq.~(\ref{eq16}) that the phenomena of chiral anomaly cancellations can be realized by assigning approximate values to $X_N$, $X_Q$, $X_u$, and $X_d$ while considering $X_E$ and $X_L$ as unconstrained. We focus on illustrating the effects of extra $U(1)_X$ local gauge group on the $W$ boson mass, hence we do not explore the chiral anomaly cancellations detailedly for different $X_E$, $X_L$ in specific NP models.

\section{Numerical analyses\label{sec3}}

Based on the calculations in Sec.~\ref{sec2}, the numerical results are computed and presented in this section. As input parameters~\cite{PDG}, $M_W^{\rm SM}$ is taken as $80.354{\rm GeV}$, the $Z$ boson mass is $M_Z=91.1876\;{\rm GeV}$, the fine-structure constant $\alpha_{em}(m_Z)=1/128.9$, and the fermion coupling constant $G_F=1.1664\times10^{-5}{\rm GeV}^{-2}$.

For $Z'$ gauge boson, the large amount of data collected at the LHC provides great potential to search $Z'$ directly. Currently, ATLAS~\cite{ATLAS:2019erb} and CMS~\cite{CMS:2021ctt} present direct searches on $Z'$ gauge boson through the channel $pp\to Z'\to e^+e^-,\;\mu^+\mu^-$ based on the data collected in proton$-$proton collisions at a centre-of-mass energy $13\;{\rm TeV}$. In addition, there is a measurement of $t\bar t$ pair production cross section with $35.9\;{\rm fb}^{-1}$ data from CMS~\cite{CMS:2020tvq} and $139\;{\rm fb}^{-1}$ data from ATLAS~\cite{ATLAS:2020aln} through the channel $pp\to Z'\to t\bar t$, and ATLAS~\cite{ATLAS:2019fgd} also present a search for new resonances decaying into a pair of jets through the channel $pp\to Z'\to q\bar q$, including $b\bar b$. Although the $Z'$ gauge boson is not observed so far, severe constraints are set on $Z'$ boson mass, which depends on the $U(1)$ symmetry and the corresponding gauge coupling strength generally. For example, for a sequential Standard Model (SSM), the latest experimental constraint on $Z'_{SSM}$ boson (which has the same fermion couplings as the SM Z boson) mass is $M_{Z'}\gtrsim5.1\;{\rm TeV}$~\cite{ATLAS:2019erb,CMS:2021ctt}. For an ${\rm E}_6-$motivated Grand Unification model, the additional gauge bosons are required to be heavier than about $4.1\;{\rm TeV}$ for $Z'_{\Psi}$ and $4.6\;{\rm TeV}$ for $Z'_{\chi}$ experimentally~\cite{ATLAS:2019erb,CMS:2021ctt}.

The new $Z'$ boson can also make contributions to $B_s-\bar B_s$ mixing and the process $B_s\to\mu^+\mu^-$, which depends on the additional $U(1)$ symmetry. For example, the contributions from $Z'$ boson to $B_s\to\mu^+\mu^-$ for the $U(1)_{B-L}$ symmetry are highly suppressed by its heavy mass~\cite{Yang:2018fvw}, while $Z'$ boson can make important contributions to $B_s-\bar B_s$ mixing and $B_s\to\mu^+\mu^-$ at the tree level for the $U(1)_{L_\mu-L_\tau}$ symmetry~\cite{Han:2019diw} and the flavor dependent $U(1)_F$ symmetry~\cite{Yang:2024kfs,Yang:2024znv}. In addition, how to probe the new introduced $Z'$ boson in future also depends on the $U(1)$ symmetry and the relevant couplings. For example, $Z'$ will be produced at the LHC if $Z'$ have nonzero couplings with quarks, which indicates the searches for high-mass dielectron, dimuon, dijet resonances at the future HL-LHC are effective to probe $Z'$ boson directly; $Z'$ has the potential to be probed by the di-muon production at the future high-energy electron-positron Linear Colliders if $Z'$ have nonzero couplings with charged leptons~\cite{Basso:2009hf}, and $Z'$ also may be produced at future muon collider~\cite{Sun:2023rsb,Martinez-Martinez:2023qjt,Gutierrez-Rodriguez:2024nny} in this case. In the following analysis, we consider the constraints on $Z'$ gauge boson mass $M_{Z'}$ and $M_{Z'}/g_X$ as~\cite{ATLAS:2016cyf,Carena:2004xs,Cacciapaglia:2006pk}
\begin{eqnarray}
&&M_{Z'}\geq4.2\;{\rm TeV},\;\frac{M_{Z'}}{g_X}\geq6\;{\rm TeV}.
\end{eqnarray}

\begin{figure}
\setlength{\unitlength}{1mm}
\centering
\includegraphics[width=1.9in]{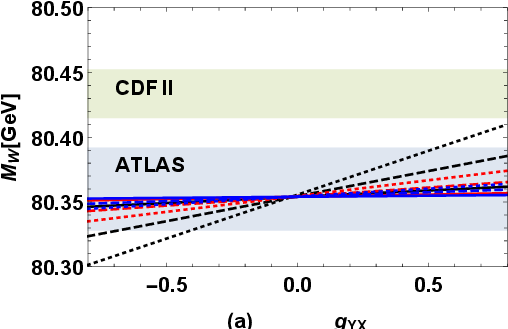}
\vspace{0.1cm}
\includegraphics[width=1.9in]{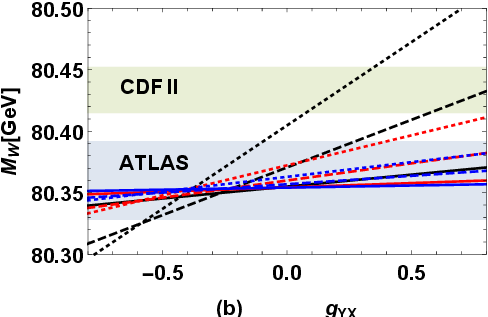}
\vspace{0.1cm}
\includegraphics[width=1.9in]{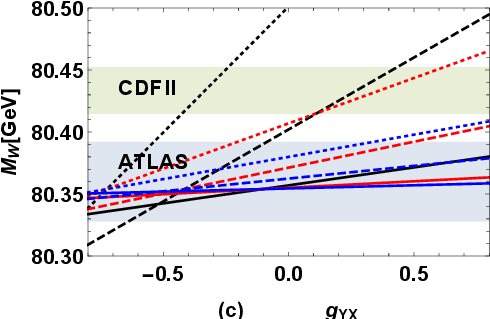}
\vspace{0cm}
\caption[]{Taking $X_E=-1/2$, $X_L=-1$, the results of $M_W$ versus $g_{YX}$ are plotted for $X_E-X_L+X_H=1/2$ (a), $1$ (b), $3/2$ (c), where the gray areas denote the ATLAS $2\sigma$ interval, the green areas denote the CDF II $2\sigma$ interval, the black, red, blue lines denote the results for $M_{Z'}=4.2,\;7,\;10\;{\rm TeV}$ respectively, and the solid, dashed, dotted lines denote the results for $g_X=0.1.\;0.4,\;0.7$ respectively.}
\label{fig1}
\end{figure}
The kinetic mixing effect appears in any NP models with two Abelian groups, we focus on the effects of $g_{YX}$ on $M_W$ with eliminating the neutral currents involving charged leptons firstly. Taking $X_E=-1/2$, $X_L=-1$, the results of $M_W$ versus $g_{YX}$ are plotted for $X_E-X_L+X_H=1/2,\;1,\;3/2$ in Fig.~\ref{fig1}~(a), (b), (c) respectively, where the gray areas denote the ATLAS $2\sigma$ interval, the green areas denote the CDF II $2\sigma$ interval, the black, red, blue lines denote the results for $M_{Z'}=4.2,\;7,\;10\;{\rm TeV}$ respectively, and the solid, dashed, dotted lines denote the results for $g_X=0.1.\;0.4,\;0.7$ respectively. The picture shows that large $g_X$, $g_{YX}$ and $X_E-X_L+X_H$ combined with small $M_{Z'}$ can well explain the measured $M_W$ at CDF II, while the one measured at ATLAS prefers small $g_X$, $g_{YX}$, $X_E-X_L+X_H$ and large $M_{Z'}$. In addition, a positive $g_{YX}$ can increase the theoretical prediction of $M_W$, while a negative $g_{YX}$ suppresses the contributions from $U(1)_X$ gauge group to $M_W$. The effects of $g_{YX}$ are affected by the values of $M_{Z'}$, $g_X$, $X_E-X_L+X_H$ obviously, where $g_{YX}$ affects $M_W$ drastically when $M_{Z'}$ is small and $g_X$, $X_E-X_L+X_H$ are large. The fact is well to be understood because the effects of introducing $U(1)_X$ local gauge group are highly suppressed by large $M_{Z'}$ or very small $g_X$, and the leptonic Yukawa couplings approach to invariant under the $U(1)_X$ symmetry as $X_E-X_L+X_H$ approaches to $0$. Since the kinetic mixing constant $g_{YX}$ may affects the theoretical predictions on $M_W$ significantly, the constraints on $M_{Z'}/g_X$ from high-precision W boson mass in the NP models with extra local $U(1)$ gauge group would be relaxed partly by considering the contributions from $g_{YX}$.

\begin{figure}
\setlength{\unitlength}{1mm}
\centering
\includegraphics[width=1.9in]{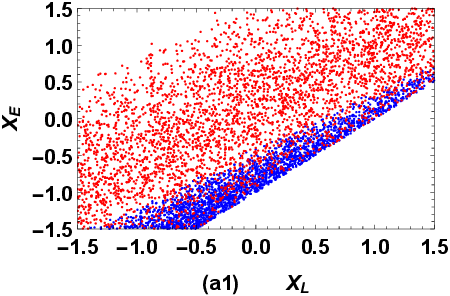}
\vspace{0.1cm}
\includegraphics[width=1.9in]{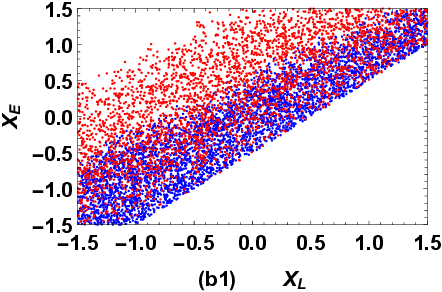}
\vspace{0.1cm}
\includegraphics[width=1.9in]{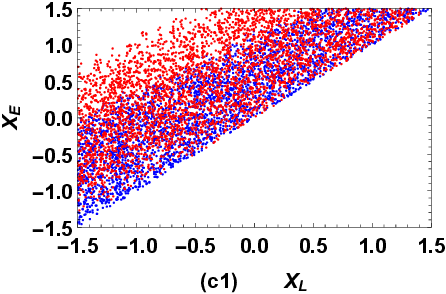}
\vspace{0cm}
\includegraphics[width=1.9in]{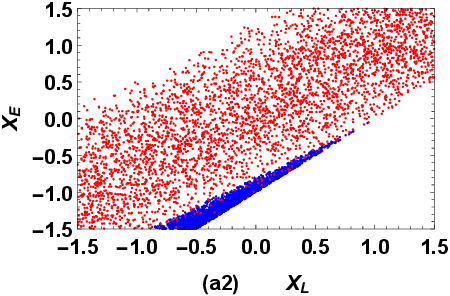}
\vspace{0.1cm}
\includegraphics[width=1.9in]{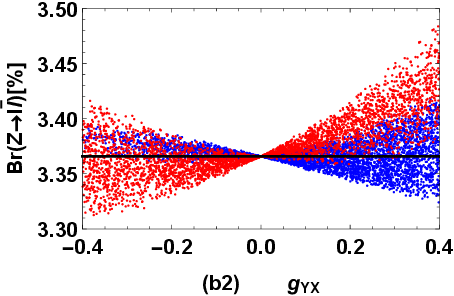}
\vspace{0.1cm}
\includegraphics[width=1.9in]{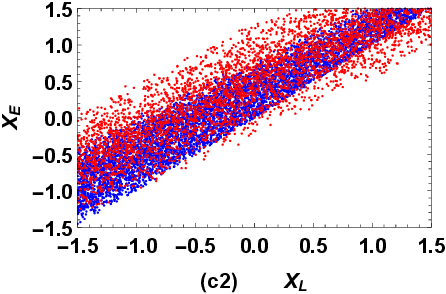}
\vspace{0cm}
\caption[]{Taking $g_B=0.6$, $M_{Z'}=5\;{\rm TeV}$, the allowed results of $X_E,\;X_L$ for $X_E-X_L+X_H=1/2$ (a1), $1$ (b1), $3/2$ (c1) are plotted, where the blue (red) points are obtained by considering the W boson mass measured at CDF II (ATLAS) in $2\sigma$ interval. Similarly, the results obtained for $g_{YX}$ scanning in the range $g_{YX}=(-0.4,0.4)$ are plotted in Fig.~\ref{fig2}~(a2), (b2), (c2).}
\label{fig2}
\end{figure}
In order to see the effects of $X_E,\;X_L,\;X_H$ on the W boson masses, we take $g_B=0.6$, $M_{Z'}=5\;{\rm TeV}$ and scan the following parameter space
\begin{eqnarray}
&&g_{YX}=(-0.8,0.8),\;X_E=(-\frac{3}{2},\frac{3}{2}),\;X_L=(-\frac{3}{2},\frac{3}{2}),\;X_H=(-\frac{3}{2},\frac{3}{2}).
\end{eqnarray}
The allowed results of $X_E,\;X_L$ for $X_E-X_L+X_H=1/2,\;1,\;3/2$ are plotted in Fig.~\ref{fig2}~(a1), (b1), (c1) respectively, where the blue points are obtained by considering the W boson mass measured at CDF II in $2\sigma$ interval, the red points are obtained by considering the W boson mass measured at ATLAS in $2\sigma$ interval. To illustrate the effects of kinetic mixing on the allowed results of $X_E,\;X_L$, the results for scanning $g_{YX}$ in the range $g_{YX}=(-0.4,0.4)$ are plotted in Fig.~\ref{fig2}~(a2), (b2), (c2) similarly. As can be seen by comparing Fig.~\ref{fig2}~(a1), (b1), (c1) with Fig.~\ref{fig2}~(a2), (b2), (c2), the allowed ranges of $X_E,\;X_L$ for $g_{YX}=(-0.4,0.4)$ are narrower than the ones for $g_{YX}=(-0.8,0.8)$, because in the chosen parameter space $g_{YX}$ affects the numerical results obviously as concluded above. For NP models with extra $U(1)$ local gauge group, the ranges of $X_E$, $X_L$ would be limited by the W boson mass measured at CDF II more strictly for smaller $X_E-X_L+X_H$, while the ranges of $X_E$, $X_L$ would be limited by the one measured at ATLAS more strictly for larger $X_E-X_L+X_H$. And the W boson mass measured at CDF II or ATLAS can be satisfied for appropriate values of $X_E$, $X_L$, $X_H$, $g_{YX}$, $g_{X}$, $M_{Z'}$.

\begin{figure}
\setlength{\unitlength}{1mm}
\centering
\includegraphics[width=1.9in]{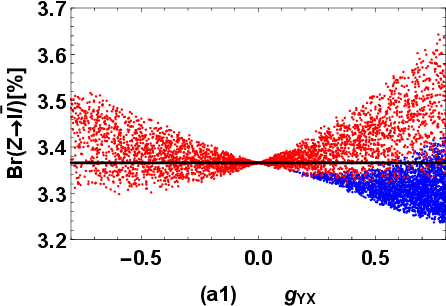}
\vspace{0.1cm}
\includegraphics[width=1.9in]{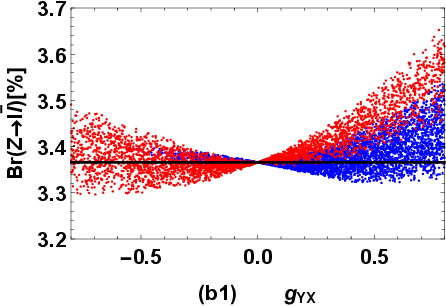}
\vspace{0.1cm}
\includegraphics[width=1.9in]{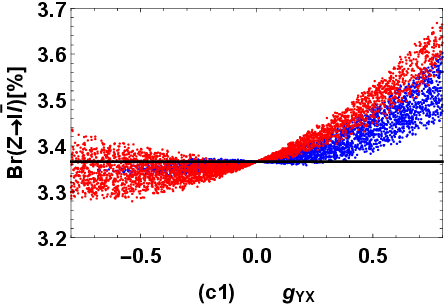}
\vspace{0cm}
\includegraphics[width=1.9in]{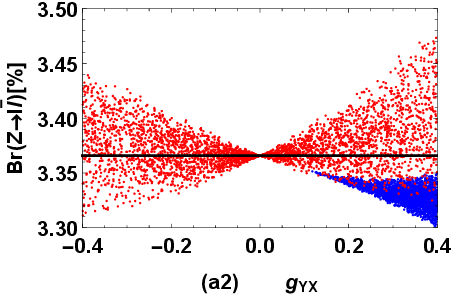}
\vspace{0.1cm}
\includegraphics[width=1.9in]{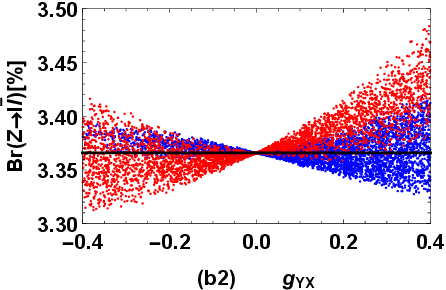}
\vspace{0.1cm}
\includegraphics[width=1.9in]{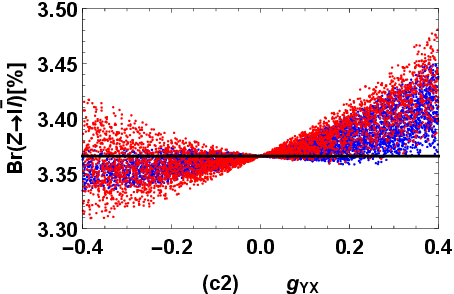}
\vspace{0cm}
\caption[]{Taking $g_B=0.6$, $M_{Z'}=5\;{\rm TeV}$, ${\rm Br}(Z\to l\bar l)$ versus $g_{YX}$ for $X_E-X_L+X_H=1/2$ (a1), $1$ (b1), $3/2$ (c1) are plotted, where the blue (red) points are obtained by considering the W boson mass measured at CDF II (ATLAS) in $2\sigma$ interval, the black lines denote the SM prediction. Similarly, the results obtained for $g_{YX}$ scanning in the range $g_{YX}=(-0.4,0.4)$ are plotted in Fig.~\ref{fig3}~(a2), (b2), (c2).}
\label{fig3}
\end{figure}
Nonzero $X_L$, $X_E$, $g_{YX}$ may affect the theoretical predictions on the branching ratio of $Z\to l\bar l$ significantly, which would be useful for testing the models with extra $U(1)$ gauge group in future experiments. Generally, ${\rm Br}(Z\to l\bar l)$ can be simplified by neglecting the final lepton masses as
\begin{eqnarray}
&&{\rm Br}(Z\to l\bar l)\approx(1+\frac{3\alpha_{em}}{4\pi})(g_V^2+g_A^2)\frac{M_Z^3}{12\pi v^2\Gamma_Z},
\end{eqnarray}
where $\Gamma_Z=2.4952\;{\rm GeV}$~\cite{PDG}, $v=246.22\;{\rm GeV}$~\cite{PDG} and
\begin{eqnarray}
&&g_V=-\frac{c_W}{2g_2}[c_W c_W' g_2-3c_W' s_W g_Y+2s_W'(g_{YX}-g_X X_E-g_X X_L)],\nonumber\\
&&g_A=-\frac{c_W}{2g_2}[c_W c_W' g_2+c_W' s_W g_Y+2s_W'g_X(X_E-X_L)],\nonumber\\
&&s_W'\approx\frac{g_{YX}M_Z}{2g_X M_{Z'}},\;c_W'=\sqrt{1-s_W'^2}.
\end{eqnarray}
To see the effects of $X_L$, $X_E$, $g_{YX}$ on ${\rm Br}(Z\to l\bar l)$ clearly, we take the points obtained in Fig.~\ref{fig2} as inputs and plot the results of ${\rm Br}(Z\to l\bar l)$ versus $g_{YX}$ in Fig.~\ref{fig3}, where (a1), (b1), (c1) present the results for $X_E-X_L+X_H=1/2$, $1$, $3/2$ respectively, the blue (red) points are obtained by considering the W boson mass measured at CDF II (ATLAS) in $2\sigma$ interval, the black lines denote the SM prediction. Fig.~\ref{fig3} (a2), (b2), (c2) present the results for $g_{YX}$ scanning in the range $g_{YX}=(-0.4,0.4)$ similarly. The picture shows obviously that $g_{YX}$ affect the numerical results of ${\rm Br}(Z\to l\bar l)$ significantly. The coupling $Z l\bar l$ in NP models with extra $U(1)$ local gauge group decouples to the one in the SM for $g_{YX}=0$ even with nonzero $X_L$, $X_E$, while ${\rm Br}(Z\to l\bar l)$ may be modified acutely for large $g_{YX}$. It indicates the kinetic mixing effect plays important roles in the $Z l\bar l$ coupling, and observing ${\rm Br}(Z\to l\bar l)$ with high precision can help to test and limit the additional gauge couplings.

\section{Summary\label{sec4}}

The contributions to the oblique parameters $S$, $T$, $U$ and W boson mass are calculated in the new physics (NP) models with extra $U(1)$ local gauge group. In such NP models, the new introduced $Z'$ gauge boson and leptonic gauge couplings can make significant contributions to $S$, $T$, $U$ and W boson mass at the tree level if we choose to eliminate the neutral currents involving charged leptons. Hence, the precise measurements of W boson mass are related closely to such NP models. Considering the kinetic mixing effects (which arises in any NP models with two Abelian groups) in the calculations, the analytical results are presented in this work, and the presented results can be applied to all such NP models. Based on the numerical results analyzed above, the kinetic mixing effects can affect the contributions to W boson mass significantly. Importantly, if the leptonic Yukawa couplings are invariant under the extra $U(1)$ local gauge group, the contributions to $S$, $T$, $U$ and W boson mass in such NP models can be eliminated by redefining the gauge boson fields through eliminating the neutral currents involving charged leptons, even with nonzero kinetic mixing effects.

\begin{acknowledgments}
The work has been supported by the National Natural Science Foundation of China (NNSFC) with Grants No. 12075074, No. 12235008, Hebei Natural Science Foundation for Distinguished Young Scholars with Grant No. A2022201017, No. A2023201041, Natural Science Foundation of Guangxi Autonomous Region with Grant No. 2022GXNSFDA035068, and the youth top-notch talent support program of the Hebei Province.

\end{acknowledgments}

\end{document}